# Effects of the Bloch-Siegert Oscillation on the Precision of Qubit Rotations: Direct Two-Level vs. Off-Resonant Raman Excitation

Prabhakar Pradhan,[1] George C. Cardoso[1], Jacob Morzinski,[2] and M.S. Shahriar[1,2]

[1]*Department of Electrical and Computer Engineering, Northwestern University Evanston, IL 60208*

[2]*Research Laboratory of Electronics, Massachusetts Institute of Technology, Cambridge, MA 02139*

In a direct two-level qubit system, when the Rabi frequency is comparable to the resonance frequency, the rotating wave approximation is not appropriate. In this case, the Rabi oscillation is accompanied by another oscillation at harmonics of twice the frequency of the driving field, the so called Bloch-Siegert oscillation (BSO), which depends on the initial phase of the driving field. This oscillation may restrict the precise rotation of a qubit made of a direct two-level system. Here, we show that in case of an effectively two-level lambda system, the BSO is inherently negligible, implying a greater precision for rotation of a qubit made of such a lambda system when compared to a direct two-level qubit in a strong driving field.

PACS Number(s): 03.67.-a, 03.67.Hk, 03.67.Lx, 32.80.Qk

# 1 . Introduction

Quantum computation [1-3] has drawn much attention in recent years due to its potential for exponentially faster computation relative to the classical case. Although the experimental realization of a quantum computer has remained a big challenge, there are several proposals for realizing them physically. It has been shown that any quantum algorithm can be decomposed into controlled-NOT gate operations and quantum bit (qubit) rotations. Qubit rotations generally make use of Rabi flopping. So far, qubit operations and toy model quantum computations have been performed by using different physical systems such as nuclear magnetic resonance, ion traps, and systems made of Josephson junctions [1-3]. In these qubit systems, the energy levels generally differ in a range of MHz to GHz. In general, a fast operation requires a high Rabi frequency which can easily be of the order of the transition frequency for these systems.

Recently we have shown that [4-6] when a two-level system is resonantly driven by a strong Rabi frequency, an effect called the Bloch-Siegert Oscillation (BSO) [7,8] becomes significant. The BSO is manifested as an oscillation of the population of either state at this frequency. The origin of the lowest order BSO for a two-level system is an effective virtual two-photon transition which occurs at an off-resonance frequency matching twice the frequency of the driving field. The magnitude of the BSO is proportional to the ratio of the Rabi frequency to the resonance frequency and also depends on the absolute phase of the driving field. The presence of the BSO changes the population of the ground and excited states as compared to the case of a weak driving field for which the rotating wave approximation (RWA) is valid. Furthermore, the dependence of the BSO on the absolute phase of the field complicates the reliable prediction of the effect of the Rabi transition on the final populations.

For a low frequency qubit system with a fast Rabi driving field, the BSO correction to the usual Rabi oscillation could be a significant fraction. For example, in a recent experiment by Martinis et al. [9], the BSO amplitude was of the order of 1% of the usual Rabi oscillation. For a still stronger driving field, this amplitude could be as large as 10%. For a fault tolerant quantum computation, the maximum permissible error rate typically scales as the inverse of the number of qubits involved in the computation. For example, a quantum computer which is made of $10^6$ qubits could tolerate well a $10^{-6}$ error rate per operation [10]. Therefore the introduction of a 1% error through an imprecise rotation of a qubit is unacceptable for most protocols.

A quantum computer operates better in a low frequency transition because of lower decoherence associated with a lower frequency. One wants qubit operations to be as fast as possible [3], so the Rabi frequency is not necessarily small [1,11,12] and may be comparable to or even greater than the transition frequency. We have shown, theoretically as well as experimentally, that [4-6] the flipping probability of the target qubit depends not only on the amplitude of the field, but on the phase of the field at location of the qubit. To avoid this potential complication, one can create a situation where the Rabi frequency is much less than the transition frequency. However, this limits the number of operations within the limited decoherence time. One solution is to keep track of the phase by measuring the phase of the field locally before each operation, but this may be technically difficult and may make the quantum computation process more complex. Here we show that the BSO effect can be avoided without limiting the operating speed if an optically off resonant Raman excitation is used to produce the Rabi flopping between two levels.

This paper is organized as follows. In section 2, we extend our previous results [4] of direct two-level qubit system for a more detailed study under a strong driving field. In section 3, we illustrate the origin of the BSO in terms of multi-photon transitions, using composite (joint) states of the quantized field and the atom. In section 4, using the composite state argument, we show that for an effective two-level lambda system where the energy difference between its two lower energy levels is much smaller than the optical frequency, the BSO is inherently negligible. We conclude with a summary in section 5.

## 2. Direct two level system (Semi-Classical Treatment)

### 2.A. General formalism for level population and BSO

We consider an ideal two-level system where a ground state $|0\rangle$ is coupled to a higher energy state $|1\rangle$. We also assume that the $|0\rangle \rightarrow |1\rangle$ transition is magnetic dipolar, with a transition frequency $\omega$ and the *classical* magnetic field is of the form $B=B_0 \cos(\omega t+\phi)$ [13]. The Hamiltonian can be written as

$$\hat{H} = \varepsilon(\hat{1}-\sigma_z)/2 + g(t)\sigma_x, \qquad (1)$$

where $g(t) = -g_0 [\exp(i\omega t+i\phi)+c.c.]/2$, $g_0$ is the Rabi frequency, $\sigma_i (i=x,y,z)$ are the standard Pauli matrices, $\varepsilon=\omega$ corresponding to resonant excitation. With the state vector

$$|\xi(t)\rangle = \begin{bmatrix} C_0(t) \\ C_1(t) \end{bmatrix}. \qquad (2)$$

Performing rotating wave transformation by a matrix

$$\hat{Q} = (\hat{1}+\sigma_z)/2 + \exp(i\omega t+i\phi).(\hat{1}-\sigma_z)/2, \qquad (3)$$

the Schrödinger equation then takes the form (setting $\hbar=1$),

$$\frac{\partial |\tilde{\xi}(t)\rangle}{\partial t} = -i\hat{\tilde{H}}(t)|\tilde{\xi}(t)\rangle. \qquad (4)$$

Now the effective Hamiltonian in the rotated basis is

$$\hat{\tilde{H}} = -g_0[1+\exp(-i2\omega t-i2\phi)]/2 \cdot \sigma_+ - g_0[1+\exp(i2\omega t+i2\phi)]/2 \cdot \sigma_-, \qquad (5)$$

and the state vector in that basis is

$$|\tilde{\xi}(t)> \equiv \hat{Q}|\xi(t)> = \begin{bmatrix} \tilde{C}_0(t) \\ \tilde{C}_1(t) \end{bmatrix}. \quad (6)$$

Writing this state vector as

$$|\tilde{\xi}(t)> = \sum_{n=-\infty}^{\infty} |\xi_n\rangle \beta^n, \quad (7)$$

where β=exp(-i2ωt-i2ϕ) and

$$|\xi_n\rangle \equiv \begin{bmatrix} a_n \\ b_n \end{bmatrix}, \quad (8)$$

one gets for all n [4]:

$$\dot{a}_n = i2n\omega a_n + ig_o(b_n + b_{n-1})/2, \quad (9a)$$

$$\dot{b}_n = i2n\omega b_n + ig_o(a_n + a_{n+1})/2. \quad (9b)$$

In figure 1, we have shown a pictorial representation of the different level interactions. In the absence of the RWA, the coupling to additional levels results from virtual multi-photon processes. Here, the coupling between $a_o$ and $b_o$ is the conventional one present when the RWA is made. The couplings to the nearest neighbors, $a_{\pm 1}$ and $b_{\pm 1}$, are detuned by an amount 2ω, and so on. Under conditions of adiabatic excitation, we get [4]

$$C_0(t) = Cos(g'_0(t)t/2) - 2\eta\Sigma \cdot Sin(g'_0(t)t/2), \quad (10a)$$

$$C_1(t) = ie^{-i(\omega t+\phi)}[Sin(g'_0(t)t/2) + 2\eta\Sigma^* \cdot Cos(g'_0(t)t/2)], \quad (10b)$$

where we have defined $g'_0(t) = \frac{1}{t}\int_0^t g_o(t')dt'$, with $g_0(t) = g_{0M}[1-\exp(-t/\tau_{sw})]$ and $\tau_{sw}$ being the switching time constant, $\eta \equiv (g_{0M}/4\omega)$ and $\Sigma \equiv (i/2)\exp[-i(2\omega t + 2\phi)]$. To lowest order in η, this solution is normalized at all times. Note that if one wants to produce this excitation on an ensemble of atoms using a π/2 pulse and measure the population of the state |1> at a time t=τ so that $g'_0(\tau)\tau/2 = \pi/2$, the result would be a signal given by

$$|C_1(g'_0(\tau),\phi)|^2 = \frac{1}{2}[1+2\eta Sin(2\omega\tau+2\phi)], \quad (11)$$

which contains information of both the amplitude and the phase of the driving field. This result shows that it is possible to determine both the phase (modulo $\pi$) and amplitude of an RF signal coupled to a two-level system by observing the population of one of these levels. A physical realization of this result can be appreciated best by considering an experimental arrangement where a thermal, effusive atomic beam is made to pass through a microwave field [4,6]. The total passage-time of an atom through the microwave field before detection is $\tau$, which includes the switching-on time. The states of the atoms are measured while the atoms are in the center of the magnetic interaction region, for example.

In reference 4, we have shown the evolution of the excited state population $|C_1(\tau)|^2$ with time $\tau$, which is the Rabi-Bloch-Siegert Oscillation (RBSO), by plotting the analytical expression in Eq. (10b). The BSO by itself is the finer rapid oscillation part of the total Rabi oscillation $|C_1(t)|^2$, i.e., $\eta \cdot \sin(g'_0(t)\,t) \cdot \sin(2\omega t + 2\phi)$ (to lowest order in $\eta$) also plotted in reference 4. These analytical results agree closely with the results we obtained via direct numerical integration of Eq. (4), provided the parameters chosen satisfy the conditions for adiabatic following.

When the value of the Rabi frequency is increased further, the finer oscillation of eqn. (4) display higher order harmonics, at multiples of $2\omega$. As such, it is useful to define the generalized BSO (GBSO), which can be identified as the variation in the population of level $|1\rangle$ from what is expected under the RWA. Figure 2, shows a plot of the GBSO amplitude, plotted as a function of the observation time, under the condition of a nominal $\pi/2$ excitation ( as assumed in the derivation of Eq. (11) ). As illustrated in reference 4, such a scenario is achievable using the atomic beam, where each atom sees the same degree of mean excitation; however, the actual time of observation varies continuously, as new atoms enter the interaction zone and then detected. As can be seen, the GBSO amplitude now shows an oscillation at both $2\omega$ as well as $4\omega$. We have observed both of these harmonics experimentally as well [6].

As discussed in the introduction, the fact that the GBSO amplitude depends on the absolute phase of the excitation field imposes constraints on precision of rotation of a quantum bit (qubit) [4]. To illustrate this explicitly, consider a generic scenario for an elementary quantum computer/register, as shown in Fig. 3. Here, we assume that the qubits are represented as simple two-level systems. The transitions between the states are assumed to be produced by Rabi flopping, induced by a microwave field. Consider operations performed on the target qubit, indicated by the arrow, assuming that it is brought to resonance with the field by somehow changing its energy levels through a scheme not relevant to our discussion here. The effective Rabi flopping will depend not only on the amplitude, but also on the phase of the field at the particular spatial point of the target qubit at the moment when the qubit interaction with the microwave field begins. This extra dependence potentially complicates the accuracy of qubit rotations employing direct two-level systems when the driving field is strong.

## 2. B. The phase of BSO for arbitrary initial populations of the two levels

For a two-level system, based on the formalism described above [4], the population amplitude and BSO can be generalized for any arbitrary initial population. Consider a situation when there is a finite population in the excited state at the beginning, i.e. $\rho_{11}(t=0) = A_0$ and $\rho_{00}(t=0) = 1 - A_0$. Now, the population amplitude for the present case can be written as

$$C_1(t) = ie^{-i(\omega t + \phi)} \{[A_0 + (1 - 2A_0)\sin^2(g_0'(t)t/2)]^{1/2} + 2\sigma\Sigma^* \cdot [A_0 + (1 - 2A_0)\cos^2(g_0'(t)t/2)]^{1/2}\}, \tag{12}$$

where $\Sigma \equiv (i/2)\exp[-i(2\omega t + 2\phi)]$.

Then, the population density of the excited state is

$$|C_1(t)|^2 = [A_0 + (1 - 2A_0)\sin^2(g_0'(t)t/2)] + (g/4\omega) \cdot (1 - 2A_0)[\sin(g_0'(t)t)]\sin(2\omega t + 2\phi). \tag{13}$$

This shows that, although the amplitude of the BSO changes with the initial population, however, the sinusoidal part of the BSO at frequency $2\omega$, i.e. $\sin(2\omega t + 2\phi)$, does not acquire any extra phase, except for a possible negative sign. This is consistent with the fact that the phase is determined in modulus $\pi$. This also is true for the BSO associated with ground state population $|C_0(t)|^2$. For $A_0 = 0$, that is, if there is no density at t=0 at the excited state, Eqs. (12) and (13) converge to the original formalism described above.

For any initial distribution of population in the two levels, the sinusoidal form of the BSO oscillation part at frequency $2\omega$, i.e. $\sin(2\omega t + 2\phi)$, does not change its form. This also can be shown by direct numerical simulation of the level density equations by arbitrary initial conditions, as shown in Fig.4. This is an important characteristics of the BSO.

## 2.C. Contribution of dc field to BSO (e.g., effect of a residual earth's magnetic field in the direction of dipole )

Consider that there is an extra non-zero value of dc magnetic field along the x direction such that the total magnetic field $\hat{B}_{total} = B_z \hat{z} + B_{dc} \hat{x} + B_0 \cos(2\omega t + 2\phi)\hat{x}$ and the Rabi frequency associated with the dc part of the magnetic field is $g_{dc}$. The equation of motion of the population amplitudes of a two-level system can be generalized by modifying the interaction part of the Hamiltonian in Eq. (5) as given below,

$$\hat{\tilde{H}}_{Total} = -g_0[1 + \frac{g_{dc}}{g_0}\exp(-i\omega t - i\phi) + \exp(-i2\omega t - i2\phi)]/2\ \sigma_+$$
$$- g_0[1 + \frac{g_{dc}}{g_0}\exp(+i\omega t + i\phi) + \exp(i2\omega t + i2\phi)]/2\ \sigma_-. \tag{14}$$

In this situation, other than the standard BSO oscillation at $2\omega$, there will be an extra oscillation at frequency $\omega$, accompanied along with the Rabi oscillation. The effect of a dc magnetic field on the BSO is described in figures 5, 6 and 7. Fig. 5 describes the change of a Rabi oscillation in the presence of dc magnetic field. Then, Fig.6 shows the BSO amplitude versus the interaction time due to the ac and dc part of the field. Finally, Fig. 7 shows the GBSO for the ac and dc part of the field. It is clear that, although dc part does not contribute to the net Rabi oscillation within RWA, it has a finite effect when the RWA is not valid. The dc part of the field contribute to net BSO amplitude in a strong driving field.

So far, we have considered the direct excitation of a two level system using a microwave field. Another way to obtain an effective two-level system in this regime, by only using light beams, is to excite an optically-off-resonant Raman transition between two low-lying states of a lambda system [17-19]. Under such an excitation, we find that the effect of BSO is negligible, even when the effective Rabi frequency of the Raman transition is much stronger than the transition frequency between the two low-lying states. In order to explain this result, it instructive first to interpret the origin of the BSO for the direct two level excitation using the quantum composite states picture, and then develop the corresponding picture for the Raman excitation. These developments are presented in the following sections.

## 3. A quantum composite state view of the BSO for a two-level atomic system under direct excitation

We assume that the quantum state of the atom is $\sum c_i |i>$, where i = 0, 1 are the ground and excited states respectively, and the quantum state of the laser field is of the form $|\Phi\rangle_n = \sum c_n |n>$, where |n> is a quantized state of $n$ photons with energy $n\omega$ ($\hbar = 1$). We will further consider that $|\Phi\rangle_n$ is a coherent state. We can then write the joint (composite) state of the laser and atoms for the two-level system as

$$|\Psi_{2L}> \equiv \sum_i c_i |i> \otimes \sum_n c_n |n>. \quad (15)$$

The atom and field interaction Hamiltonian, without RWA, can be written as:

$$\hat{H}_{I(2L)} = g(\hat{S}_{01} + \hat{S}_{10}) \otimes (a + a^+), \quad (16)$$

where $g$ is the Rabi frequency for the transition 0→1, assumed to be real, $a^+$ and $a$ are the creation and the annihilation operators, respectively, of the laser field, and $\hat{S}_{ij}$ (i,j=0,1) is the atomic level transition operator given by $\hat{S}_{ij} = |i\rangle\langle j|$.

Now the matrix element due to the 2-level atom field (2L) interaction Hamiltonian is given by

$$< H_{I(2L)} > = \langle \Psi_{2L} | H_{I(2L)} | \Psi_{2L} \rangle$$

$$= g \sum_{j=0,1} c_j^* \sum_{i=0,1} c_i \sum_{n'=-\infty,+\infty} c_{n'}^* \sum_{n=-\infty,+\infty} c_n \langle j | \langle n' | H_{I(2L)} | n \rangle | i \rangle$$

$$= g \sum_{j=0,1} c_j^* \sum_{i=0,1} c_i \sum_{n'=-\infty,+\infty} c_{n'}^* \sum_{n=-\infty,+\infty} c_n (\delta_{j,1}\delta_{i,0} + \delta_{j,0}\delta_{i,1}) \cdot (\delta_{n',n-1}\sqrt{n} + \delta_{n',n+1}\sqrt{n+1}). \quad (17)$$

The interaction is illustrated in Fig. 8. The first set of permitted transitions between the two composite states are at zero detuning when the composite states are in the same energy, and these come from the type of transitions $|n\rangle|i=0\rangle \leftrightarrow |n-1\rangle|i=1\rangle$. The second set of allowed transitions are at a detuned frequency $2\omega$, and are associated with the type of transitions $|n\rangle|i=0\rangle \leftrightarrow |n+1\rangle|i=1\rangle$. This is the interaction that leads to the BSO. Other allowed transitions which are detuned by $4\omega$, $6\omega$, etc, are also possible; however, amplitude of these processes are increasingly weaker. The GBSO corresponds to all the off resonant transitions of this type.

## 4. A Raman transition in a 2-level Lambda system and BSO

In a Raman system, states |0> and |2> are the two low-lying nearby states, and a third level |1> is at an optical frequency away from the levels |0> and |2>. Now, we apply two off-resonant driving fields at frequencies $\omega_{01} + \delta$ for the $|0\rangle \leftrightarrow |1\rangle$ transition and $\omega_{12} + \delta$ for the $|1\rangle \leftrightarrow |2\rangle$ transition, where $\delta$ is the detuning frequency. For simplicity, we consider that the optical Rabi frequencies for the interactions $|0\rangle \leftrightarrow |1\rangle$ and $|1\rangle \leftrightarrow |2\rangle$ are same and of magnitude g, $\omega_{01} \approx \omega_{12} = \omega$ and $\Delta\omega - \omega_{01} = \omega_{02}$. These assumptions result in an effective Rabi oscillation between the levels |0> and |1> at the Raman Rabi frequency of $\Omega_{02} \approx g^2/4\delta$ for $\delta \gg g$. We want to derive what is the effective BSO between 0-2 which are differ by frequency $\Delta\omega$.

The 3-level atom and fields (3L) interaction Hamiltonian, without RWA, can be written as the sum of the interaction of atom and two separate field modes

$$\hat{H}_{I(3L)} = \hat{H}_{I_1(3L)} + \hat{H}_{I_2(3L)}$$

$$= [g\ (S_{01} + S_{10}) \otimes (a_1 + a_1^+)] + [g\ (S_{12} + S_{21}) \otimes (a_2 + a_2^+)], \quad (18)$$

where g is the Rabi frequency for both the transitions, $0 \rightarrow 1$ and $1 \rightarrow 2$, $a_1^+$ and $a_1$ are creation and annihilation operators, respectively, for $0 \rightarrow 1$ transition laser field (mode 1) and $a_2^+$ and $a_2$ are the corresponding operators for $1 \rightarrow 2$ transition laser field (mode 2). $S_{ij}$ (i,j=0,1,2) is the atomic level projection operator given by $S_{ij} = |i\rangle\langle j|$.

Similarly, following the described 2-level system, a joint (composite) state of the 2 laser modes with 3-level atom can be written as

$$|\Psi_{3L}> \equiv [\sum_{i_2=0,1,2} c_i |i> \otimes \sum_{m=-\infty,+\infty} c_m |m\rangle \otimes \sum_{n=-\infty,+\infty} c_n |n\rangle] \tag{19}$$

Now the transition matrix element due to the atom-field interaction is

$$<H_{I(3L)}> = \langle \Psi_{3L} | H_{I(3L)} | \Psi_{3L} \rangle$$

$$= g \sum_{j=0,1,2} c_j^* \sum_{i=0,1,2} c_i \sum_{m'=-\infty,+\infty} c_{m'}^* \sum_{m=-\infty,+\infty} c_m \sum_{n'=-\infty,+\infty} c_{n'}^* \sum_{n=-\infty,+\infty} c_n \langle j|\langle m'|\langle n'| H_{I_1(3L)} |n\rangle|m\rangle|i\rangle$$

$$+ g \sum_{j=0,1,2} c_j^* \sum_{i=0,1,2} c_i \sum_{m'=-\infty,+\infty} c_{m'}^* \sum_{m=-\infty,+\infty} c_m \sum_{n'=-\infty,+\infty} c_{n'}^* \sum_{n=-\infty,+\infty} c_n \langle j|\langle m'|\langle n'| H_{I_2(3L)} |n\rangle|m\rangle|i\rangle.$$

$$= g \sum_{j=0,1,2} c_j^* \sum_{i=0,1,2} c_i \sum_{n'=-\infty,+\infty} c_{n'}^* \sum_{n=-\infty,+\infty} c_n (\delta_{j,1}\delta_{i,0} + \delta_{j,0}\delta_{i,1}).(\delta_{n',n-1}\sqrt{n} + \delta_{n',n+1}\sqrt{n+1}).\delta_{m',m}$$

$$+ g \sum_{j=0,1,2} c_j^* \sum_{i=0,1,2} c_i \sum_{m'=-\infty,+\infty} c_{m'}^* \sum_{m=-\infty,+\infty} c_m (\delta_{j,2}\delta_{i,1} + \delta_{j,1}\delta_{i,2}).(\delta_{m',m-1}\sqrt{m} + \delta_{m',m+1}\sqrt{m+1}).\delta_{n',n}. \tag{20}$$

The interaction on each leg of the Raman transition can be pictured in the same way as in Fig. 8. The composite state picture of the Raman interaction is illustrated in Fig. 9. Any BSO which may occur between $|0>$ and $|2>$ ($\Delta\omega = \omega_{01} - \omega_{12}$) has to go through at least one two photon transition detuned by a frequency $2\omega$. Therefore, the amplitude of the BSO is of the order of $(g/\omega)$ in this lambda system, and not of the order of $\Omega_{02}/4\Delta\omega$, as in the direct two-level case, where $\Omega_{02}$ is the effective Rabi frequency between the states $|0>$ and $|2>$. In the case when $\Delta\omega$ is in microwave regime and $\omega$ is in optical regime, the $\Omega_{02}/\Delta\omega$ (or $\Omega_{02}/\omega$) is very much smaller than one, and this makes the BSO amplitude negligible in an effectively two-level lambda system.

## 5. Conclusions

We have analyzed the origin of the Bloch-Siegert shift and oscillation caused by strong fields. For a direct two-level qubit, we have shown that the presence of the Bloch-Siegert oscillation implies that the final state of a qubit rotation is a function of the absolute local phase of the driving field. This complicates the use of strong fields in two-level systems, though faster driving fields are required for faster qubit rotation. We also analyzed the case in which a qubit is formed from an effectively two-level lambda system and the rotation is performed via an optical Raman transition. In this case, we show that the BSO is negligible even when the effective Rabi frequency for the qubit rotation exceeds the transition frequency. We conclude that qubits formed by an optically Raman excited microwave transition (lambda system) may be more controllably rotated than direct microwave two-level qubits. Thereby, the qubits which are made of lambda systems may be better suited than the qubits that are made of direct two level systems for quantum computation and quantum information processing.

This work was supported by DARPA grant No. F30602-01-2-0546 under the QUIST program, ARO grant No. DAAD19-001-0177 under the MURI program, and NRO grant No. NRO-000-00-C-0158.

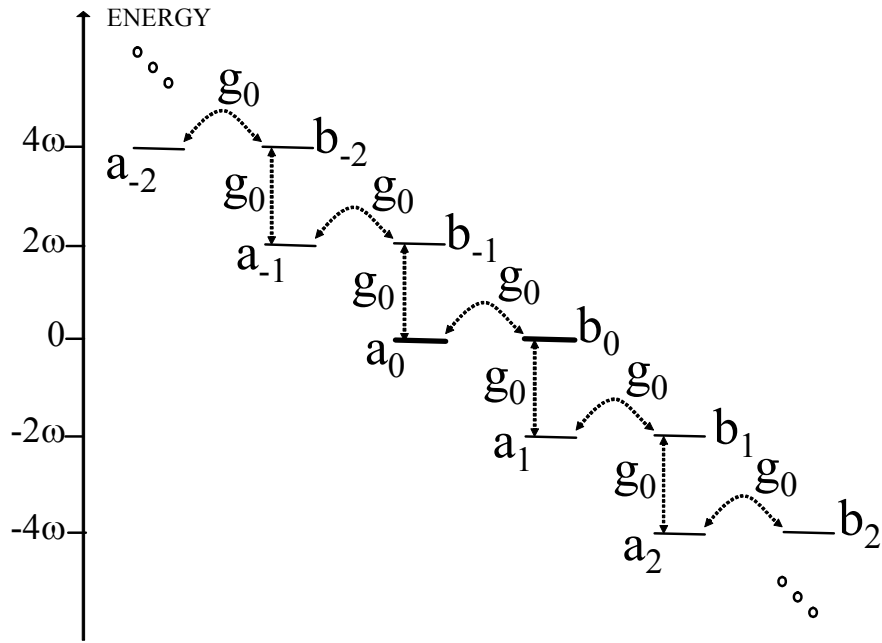

FIG. 1. Schematic illustration of the multiple orders of interaction when the rotating wave approximation is not made. The strengths of the first higher order interaction, for example, is weaker than the zeroth order interaction by the ratio of the Rabi frequency, $g_o$, and the effective detuning, $2\omega$. When the RWA is made, only the terms $a_0$ and $b_0$ may be non-zero.

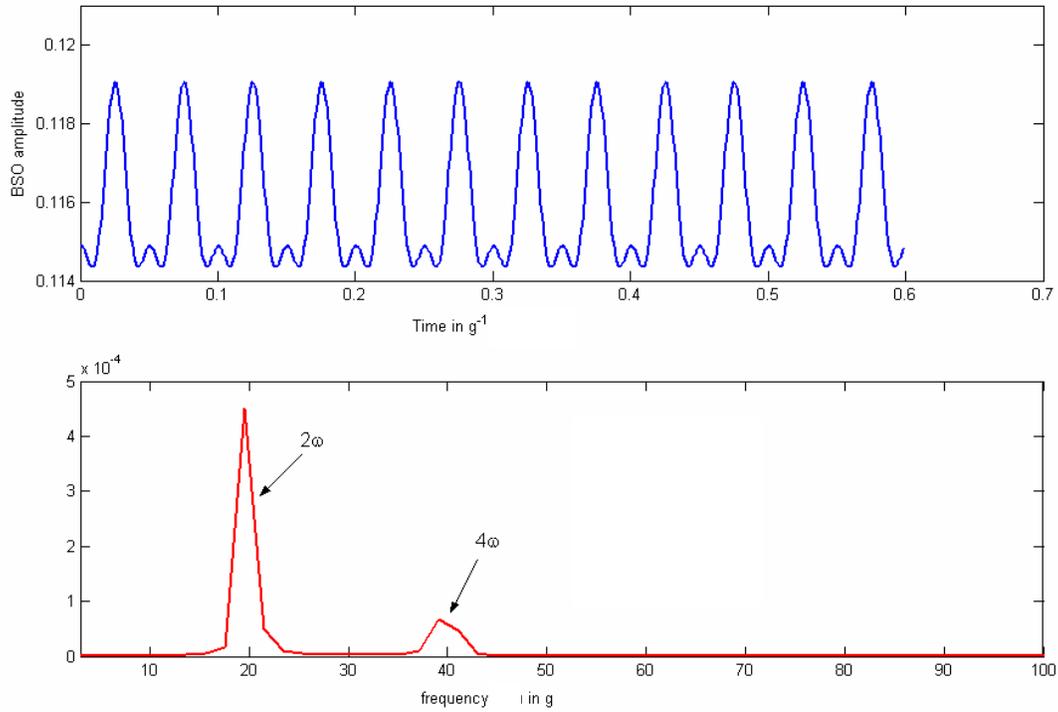

FIG. 2. (Top) Amplitude of the generalized BSO(GBSO), plotted as a function of the observation time, while the degree of excitation is kept constant, corresponding to a nominal $\pi/2$ excitation under RWA. Because of a larger Rabi frequency, the GBSO amplitude shows a visible $2\omega$ and $4\omega$ peak. (Bottom) The Fourier transform of the GBSO amplitude shows the presence of two prominent frequencies $2\omega$ and $4\omega$ where $\omega$ is the frequency of the driving field.

14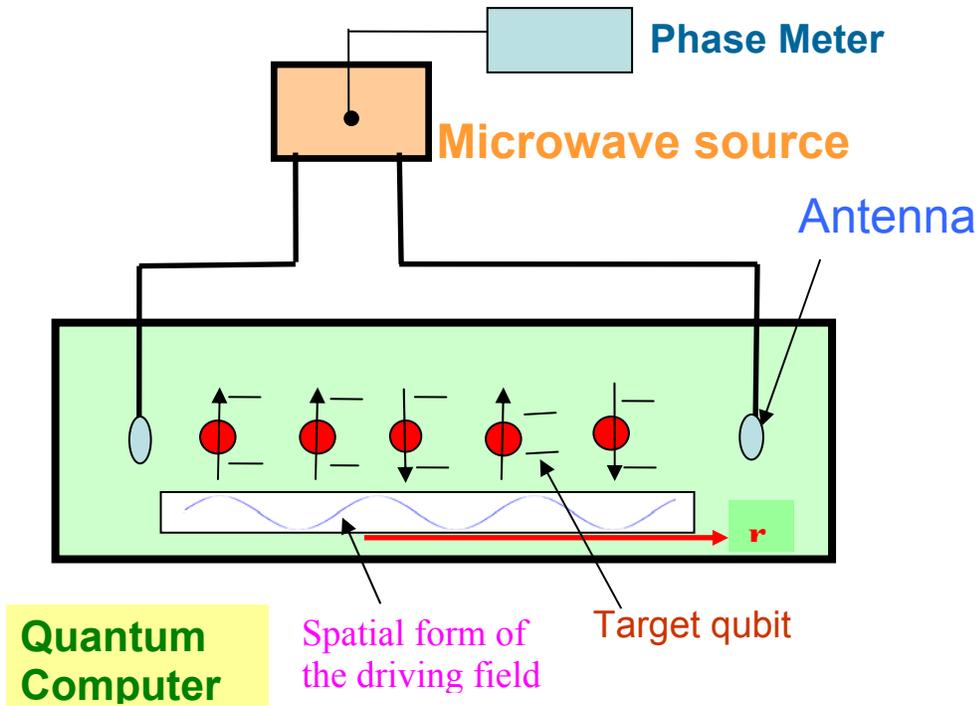

FIG. 3. In this schematic picture we have shown fixed qubits (spins) and their energy levels inside a generalized, model quantum computer / register. The target qubit is shown by an arrow. The microwave field that interact with the target qubit is shown by a sinusoidal curve varying spatially in space. The BSO effect makes the transition probability depend not only on the amplitude, but also on the phase of the microwave field at the spatial point of the target qubit.

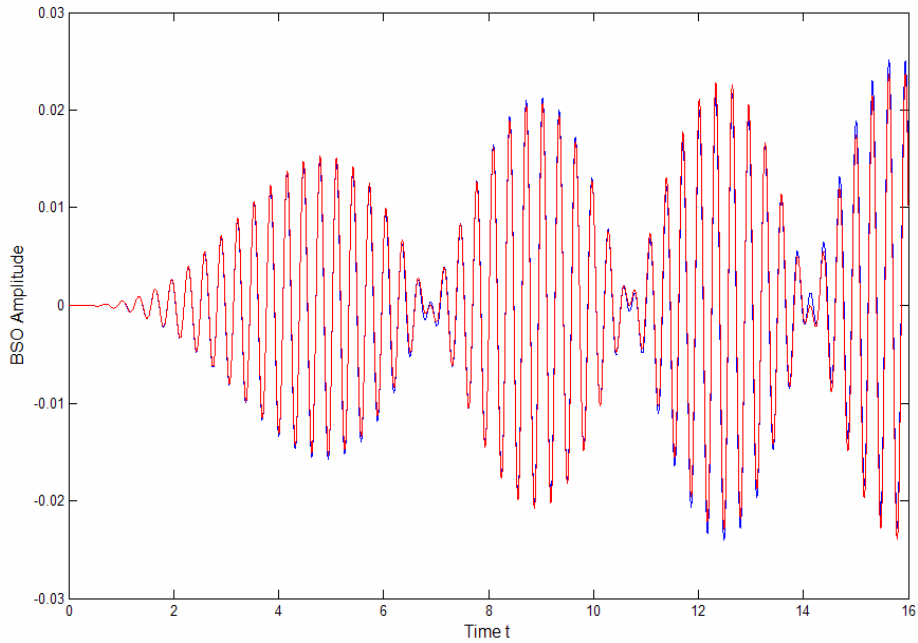

Fig.4. BSO amplitudes vs. interaction time t show a sinusoidal curve for arbitrary initial population. The initial populations are: $\rho_{00}(t=0)=.7$ and $\rho_{11}(t=0)=.3$, resonant frequency $\omega=10$, and the Rabi frequency $g=1$. The numerical simulation of the BSO amplitude (blue) is plotted together with the theoretical formula $(g/4\omega)\cdot(1-2A_0).[\sin(g'_0(t)t)].\sin(2\omega t)$, here $\phi=0$. The plots matches quite well. This shows that the form of BSO at frequency $2\omega$ is $\sin(2\omega t)$ and is independent of the initial population.

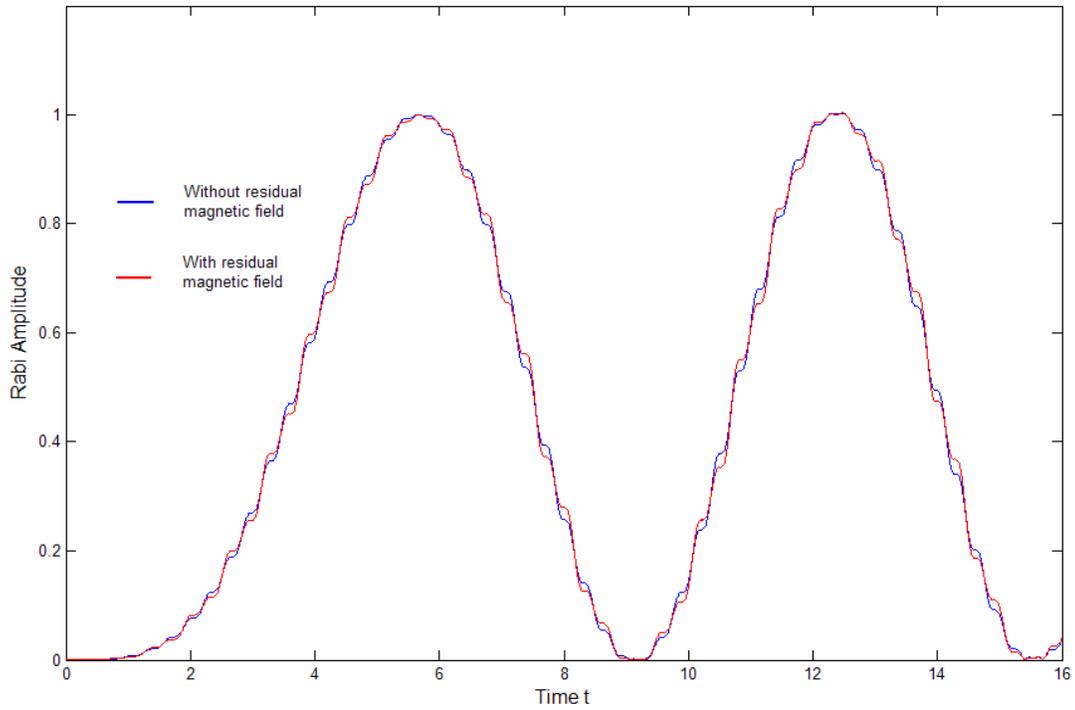

Fig. 5: Rabi oscillation in the absence (blue line) and presence (red line) of a dc magnetic field along the $\hat{x}$ direction. The simulation parameters are: initial population density $\rho_{00}(t=0)=1$, resonant frequency $\omega=10$, and the Rabi frequency $g=1$ *and* $g_{dc}=.5$.

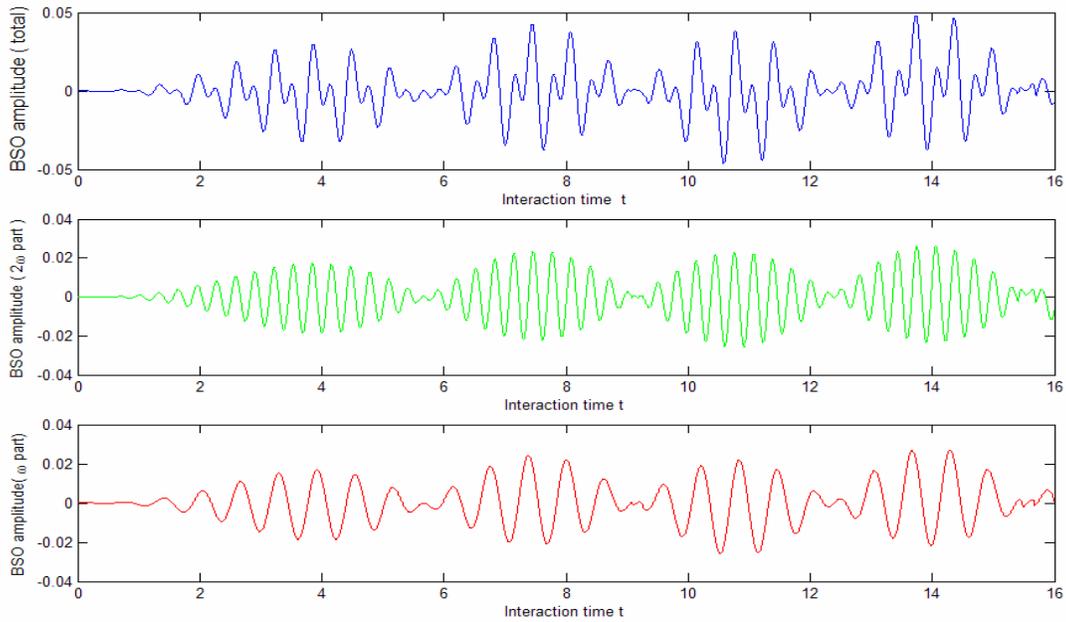

Fig. 5: BSO amplitude in presence of dc magnetic field versus atom-field interaction time. The parameters are same as Fig.4. (a) Total BSO amplitude versus interaction time t (blue line), (b) BSO amplitude associate with $2\omega$ part of the frequency versus interaction time t (green line). (c) BSO amplitude (red line) associated with $\omega$ part of the frequency versus interaction time. This is to be noted that an extra oscillation at frequency $\omega$ is present along with the standard BSO oscillation at frequency $2\omega$ due to the presence of a dc magnetic field.

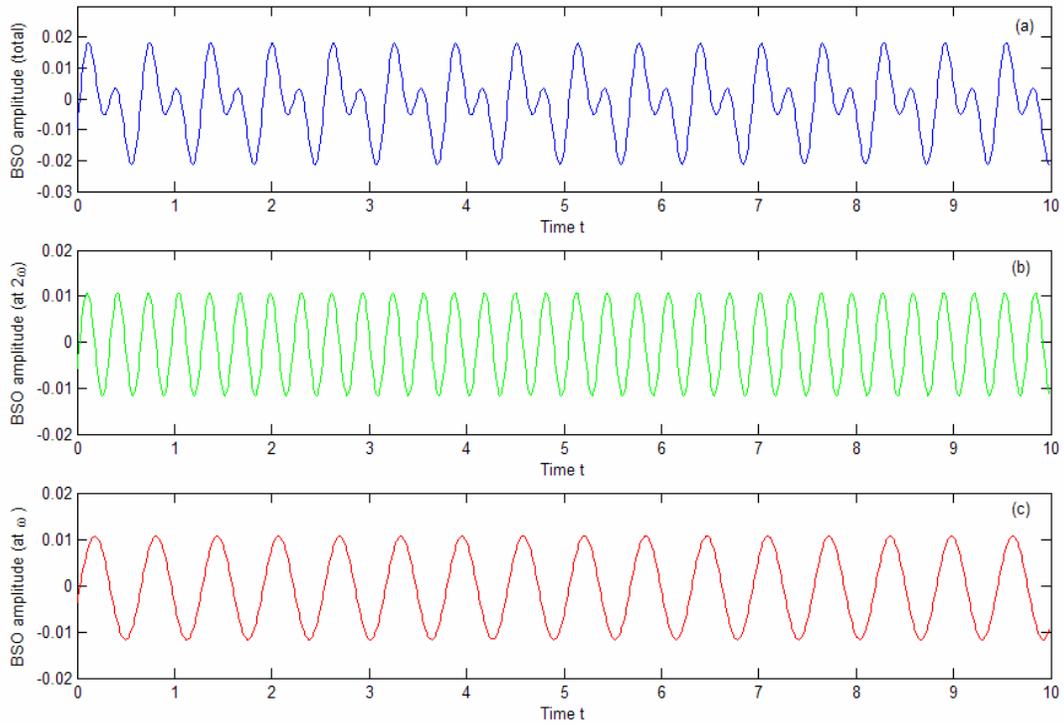

Fig. 7: GBSO amplitude versus time in the presence of an ac and dc magnetic field when the interaction time is fixed ($\tau = 5$) and magnetic field is evolving with time. GBSO is calculated at the end of the fixed interaction time, and the GBSO is varying mainly with the temporal phase of the incoming magnetic field at time $\tau + t$. Rest of the parameters are same as Fig.4. (a) Total GBSO amplitude versus time t (blue line), (b) BSO amplitude at frequency $2\omega$ versus time t (green line). (c) GBSO amplitude (red line) at frequency $\omega$ versus time. Total BSO amplitude is sum of the GBSO at $2\omega$ plus at $\omega$.

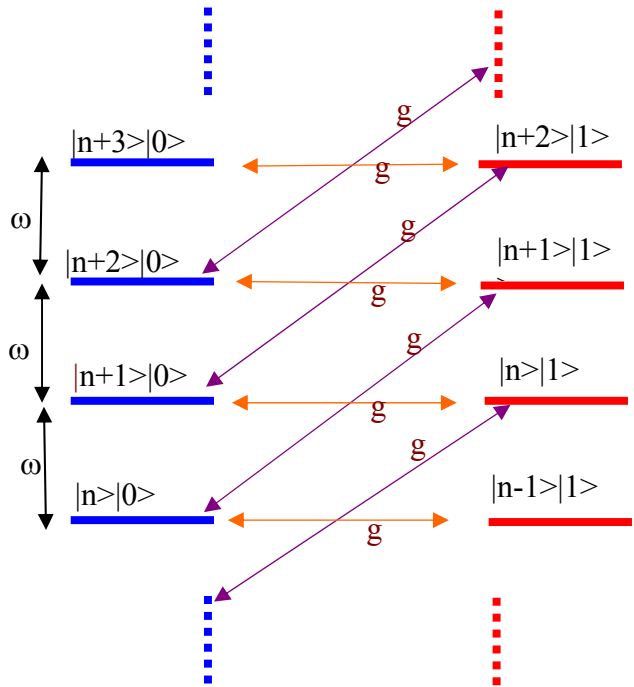

FIG. 8. Joint states of atoms and laser of a two level system and their allowed transitions are illustrated. A two level system (|0> and |1>) of resonance frequency ω, is interacting with a laser field and g is the 0-1 Rabi transition frequency. The joint states in the left side of the figure shows infinite manifolds of laser field with the ground state |0>, i.e. |n>|0> (n= …n-1, n, n+1,…), and the subsequent states are separated by an energy ω. Similarly manifolds of the excited state|1> with the laser field, i.e. |n-1>|1> (n= …n-1, n, n+1,…), are shown in the right side of the figure. Same energy states are kept along the same horizontal line, e.g., |n>|0> and |n-1>|1> states are the same energy states. The arrows indicate the allowed transitions according to Eq. (19). The transitions indicated by the horizontal arrows (orange) are due to one photon joint state transitions with zero detuning and the rate of transition is the Rabi frequency g. The inclined arrows (purple) indicate the allowed transitions that are detuned by a frequency 2ω and are responsible for the Bloch-Siegert oscillation (BSO), and each transition rate also is at a frequency g.

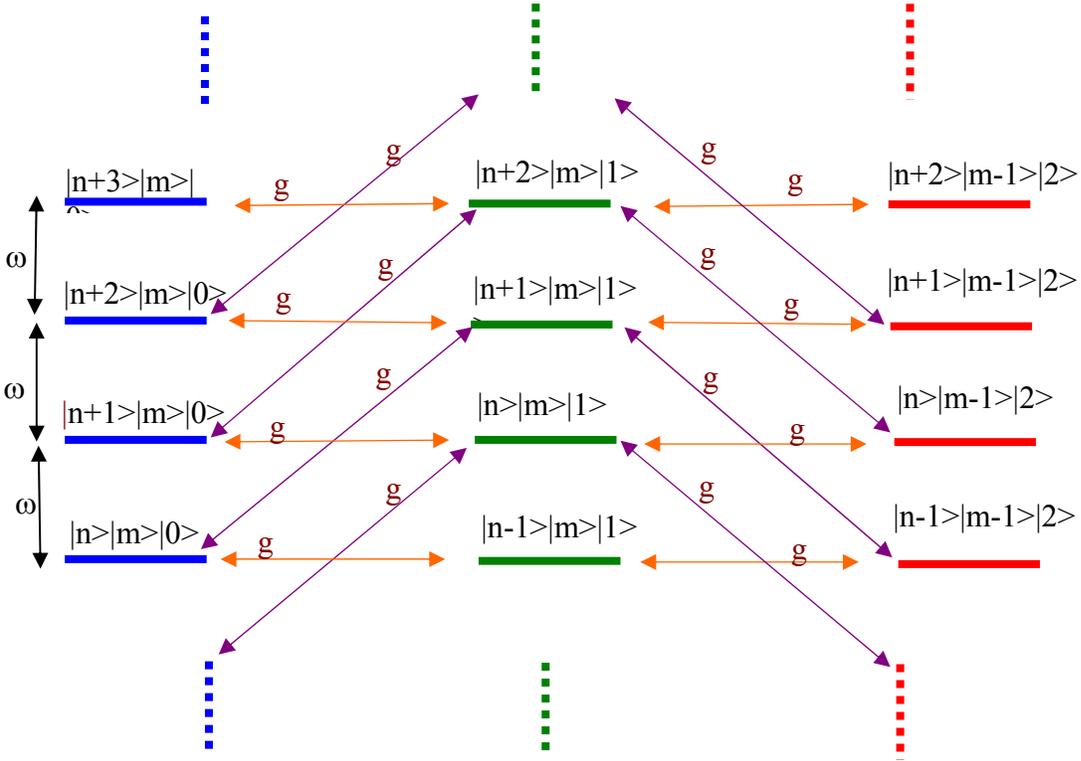

FIG. 9. Schematic illustration of a Lambda system with laser fields and their transitions. Let, g is the Rabi frequency for both the transitions, $0-1$ transition by the driving field mode 1 ( $|n\rangle$ ) and for $1-2$ transition by field mode 2 ($|m\rangle$). For simplicity, we assume that frequencies of the laser fields are same as $(\omega_{01} \approx \omega_{12} = \omega)$, $(\Delta\omega = \omega_{12} - \omega_{01})$, detuning $\delta$ is small and $\ll \omega$. (In this picture we have assumed $\delta = 0$.) Energy levels of the atom plus fields composite states, $|n\rangle|m\rangle|i\rangle$, ( where m = ..,m-1, m, m+1,…, n = .., n-1, n, n+1,… and, i = 0, 1, 2 ) are shown in three columns for $|0\rangle$, $|1\rangle$, and $|2\rangle$ states. The composite states, $|n\rangle|m\rangle|i\rangle$, have different manifolds and allowed transitions. We present here one of the possible and easiest transition path from $|0\rangle$ to $|2\rangle$ via $|1\rangle$, in search of an effective BSO for 0-2 transition. Same energy composite states are kept along the same line horizontally. The allowed transitions are according to Eq. (20). The horizontal arrows (orange) represent the resonant one-photon transitions responsible for the usual Rabi flopping, detuned at a frequency 0. The inclined arrows (purple) represent non-resonant two-photon transitions, and the transitions virtually detuned by $2\omega$, responsible for the BSO. It can be noted that there is no direct BSO effect for transition $|0\rangle - |2\rangle$, but via state $|1\rangle$. This implies, BSO for any $|0\rangle-|2\rangle$ transition will be associated with at least a $2\omega$ detuned 2-photon transition. For a lambda system, $\omega$ is in optical frequency region. This makes an effective BSO amplitude is effectively zero, because BSO amplitude is $g/(4\omega)$.